# Observation of γ-vibrations and alignments built on non-ground-state configurations in $^{156}$Dy

S. N. T. Majola[1,2]*, D. J. Hartley[3], L.L. Riedinger[4], J.F. Sharpey-Schafer[5], J. M. Allmond[6,17], C. Beausang[6], M.P. Carpenter[7], C.J. Chiara[8], N. Cooper[9], D. Curien[10], B. J. P. Gall[10], P.E. Garrett[11], R. V. F. Janssens[7], F.G. Kondev[8], W.D. Kulp[12], T. Lauritsen[7], E.A. McCutchan[7,13], D. Miller[4], J. Piot[10], N. Redon[14], M.A. Riley,[15] J. Simpson[16], I. Stefanescu[7], V. Werner[9], X. Wang[15], J.L. Wood[12], C.-H.Yu[17] and S. Zhu[7]

*[1] iThemba LABS, National Research Foundation, P. O. Box 722, Somerset-West 7129, South Africa*
*[2]University of Cape Town, Department of Physics, P/B X3, Rondebosch 7701, South Africa*
*[3]Department of Physics, US Naval Academy, Annapolis, MD 21402, USA*
*[4]Department of Physics, University of Tennessee, Knoxville, TN 37996, USA*
*[5]University of the Western Cape, Department of Physics, P/B X17, Bellville 7535, South Africa*
*[6]Department of Physics, University of Richmond, Richmond, VA 23173, USA*
*[7]Physics Division, Argonne National Laboratory, Argonne, IL 60439, USA*
*[8]Nuclear Engineering Division, Argonne National Laboratory, Argonne, IL 60439, USA*
*[9]Wright Nuclear Structure Laboratory, Yale University. New Haven, CT 06520, USA*
*[10]Université de Strasbourg, IPHC, 23 rue du Loess 67037 Strasbourg, France*
*CNRS, UMR7178, 67037 Strasbourg, France*
*[11]Department of Physics, University of Guelph, Guelph, Ontario, Canada*
*[12]School of Physics, Georgia Institute of Technology, Atlanta, GA 30332, USA*
*[13] National Nuclear Data Centre, Brookhaven National Laboratory, Upton, NY 11973, USA*
*[14] Institut de Physique Nucléaire Lyon, IN2P3-CNRS, Lyon, F-69622 Villeurbanne, France*
*[15]Department of Physics, Florida State University, Tallahassee, FL 32306*
*[16]STFC Daresbury Laboratory, Daresbury, Warrington, WA4 4AD, UK*
*[17]Physics Division, Oak Ridge National Laboratory, Oak Ridge, TN 37831*

**email address; majola@tlabs.ac.za*

The exact nature of the lowest $K^{\pi}=2^+$ rotational bands in all deformed nuclei remains obscure. Traditionally they are assumed to be collective vibrations of the nuclear shape in the γ degree of freedom perpendicular to the nuclear symmetry axis. Very few such γ-bands have been traced past the usual back-bending rotational alignments of high-*j* nucleons. We have investigated the structure of positive-parity bands in the *N=90* nucleus $^{156}$Dy, using the $^{148}$Nd($^{12}$C,4n)$^{156}$Dy reaction at 65 MeV, observing the resulting γ-ray transitions with the Gammasphere array. The even- and odd-spin members of the $K^{\pi}=2^+$ γ-band are observed to $32^+$ and $31^+$ respectively. This rotational band faithfully tracks the ground-state configuration to the highest spins. The members of a possible γ-vibration built on the aligned yrast S-band are observed to spins $28^+$ and $27^+$. An even-spin positive-parity band, observed to spin $24^+$, is a candidate for an aligned S-band built on the seniority-zero configuration of the $0_2^+$ state at 676 keV. The crossing of this band with the $0_2^+$ band is at $\hbar\omega_c$= 0.28(1) MeV and is consistent with the configuration of the $0_2^+$ band not producing any blocking of the monopole pairing.

## I. INTRODUCTION

The γ degree of freedom, breaking the axial symmetry in quadrupole deformed nuclei, is indispensable in understanding the level structures observed in nuclear spectroscopy. The traditional interpretation [1] of the lowest $K^{\pi}=2^+$ rotational bands, that consistently appear in even-even deformed nuclei, has been that they are a shape vibration in the γ degree of freedom perpendicular to the symmetry axis. These bands have both even (natural parity) and odd spin (unnatural parity) states and are observed to have very weak M1 ΔJ=1 transitions between the even and odd members of the bands [2]. The signature splitting between the odd and even spins depends on the nature of the total energy dependence of the nucleus on the shape parameter γ.

Extreme scenarios of this dependence are (a) γ-rigid and finite, as calculated by Davydov and Filippov [3] and (b) γ-soft, calculated by Wilets and Jean [4].

In odd-A nuclei, the single nucleon in a Nilsson orbital $[N,n_z,\Lambda]\Omega$ can couple to the collective motion of the even-even core. When coupling to the $K^{\pi}=2^+$ excitations, two rotational bands will result with the $K$ quantum numbers (angular momentum projection on the symmetry axis) coupled either parallel or anti-parallel; thus $K_> = (\Omega+2)$ and $K_< = |\Omega-2|$. Examples of such couplings have been observed in $^{165}$Ho and $^{167}$Er [5] and in $^{155}$Gd [6,7]. A notable feature of **γ**-bands is that they track the intrinsic configuration, usually the ground-state, on which they are based [8, 9].

To date there are few examples of γ-bands extended to high spins. This is, no doubt, due to these γ-bands lying about 1 MeV above the yrast line in most nuclei and therefore only weakly populated in most fusion-evaporation reactions. In Table I we list examples of even-even nuclei in which the γ-band has been observed to spins higher than 15 ħ.

To our knowledge, the first example of a γ-vibration, built on an $i_{13/2}^2$ aligned neutron S-band, that was observed beyond the first back-bend was in $^{164}$Er [10,11]. In $^{156}$Er, the positive-parity members of the γ-band are known through the back-bend up to 26 ħ [12]. In $^{160}$Er, the odd-spin members of the γ-band are observed through both the $i_{13/2}^2$ neutron alignment and the $h_{11/2}^2$ proton alignment to spin 43 ħ [9]. It appears that, regardless of the underlying microscopic configurations involved in the **γ**-bands, they do not seem to be affected by the orbitals responsible for causing the alignments.

**Table 1.** *Some even-even nuclei in which the γ-band has been observed above spin $15^+$.*

| *Nucleus* | *Beam* | | *Highest spin reached* | | | *Reference* |
|---|---|---|---|---|---|---|
| | *Species* | *Energy (MeV)* | *Yrast band* | *γ-even* | *γ-odd* | |
| $^{104}$Mo[a] | ff | | $20^+$ | $18^+$ | $17^+$ | [13] |
| $^{154}$Gd | α | 45 | $24^+$ | $16^+$ | $17^+$ | [14] |
| $^{156}$Dy | $^{12}$C | 65 | $32^+$ | $32^+$ | $31^+$ | this work |
| $^{160}$Dy | $^{7}$Li | 56 | $28^+$ | $22^+$ | $25^+$ | [15] |
| $^{162}$Dy[b] | $^{118}$Sn | 780 Coulex | $24^+$ | $18^+$ | $17^+$ | [16] |
| $^{162}$Dy | $^{7}$Li | 56 | $28^+$ | $18^+$ | $17^+$ | [15] |
| $^{164}$Dy[b] | $^{118}$Sn | 780 Coulex | $22^+$ | $18^+$ | $11^+$ | [16] |
| $^{156}$Er | $^{48}$Ca | 215 | $26^+$ | $26^+$ | $15^+$ | [12] |
| $^{160}$Er | $^{48}$Ca | 215 | $50^+$ | - | $43^+$ | [9] |
| $^{164}$Er | $^{9}$Be | 59 | $24^+$ | $14^+$ | $19^+$ | [17] |
| $^{164}$Er | $^{18}$O | 70 | $24^+$ | $18^+$ | $21^+$ | [10,11] |
| $^{170}$Er | $^{238}$U | 1358 Coulex | $26^+$ | $18^+$ | $19^+$ | [18] |
| $^{180}$Hf | $^{136}$Xe | 750 Coulex | $18^+$ | $16^+$ | $13^+$ | [19] |
| $^{238}$U | $^{209}$Bi | 1130 &1330 Coulex | $30^+$ | $26^+$ | $27^+$ | [20] |

[a]fission fragment. [b]inverse reaction

Early microscopic models of collective vibrations in deformed nuclei [21-25] assumed the existence of a vibrational "phonon" or "boson" and then constructed this entity out of some set of basis states. This is accomplished by postulating an interaction, expanding the collective phonon in a truncated basis, and then using the variational principle to minimize the phonon energy in terms of the interaction parameters. An extensive literature exists that discusses optimization of bases, truncations and fitting the parameters. Phenomenological models, such as the IBA [26] and geometric approaches [27], suffer from the disadvantage that they say nothing about the underlying microscopic configurations, unlike RPA calculations that appear to provide a better description [28-31]. The recent Triaxial Projected Shell Model (TPSM) calculations [32-36] seem to give a real hope of obtaining a clear microscopic and physically accurate picture of **γ**-vibrations of deformed nuclei. Reference [35] provides a clear history of approaches to descriptions of **γ**-vibrations.

In this paper, we investigate the positive-parity medium-spin structure of $^{156}$Dy and interpret the data in terms of aligned bands and γ-vibrations.

## II. EXPERIMENTAL DETAILS AND ANALYSIS

We have used the $^{148}$Nd ($^{12}$C, 4n) $^{156}$Dy reaction with a 65 MeV $^{12}$C beam from the Argonne National Laboratory ATLAS accelerator facility to populate excited states in $^{156}$Dy. The target was 1.5 mg cm$^{-2}$ of $^{148}$Nd with a 1.6 mg cm$^{-2}$ layer of natural Pb evaporated on the back and a 1.5 μg cm$^{-2}$ flash of Au evaporated on the front surface. The Pb layer was to ensure that products from fusion-evaporation reactions stopped in the target. The Au layer was to reduce oxidation of the target while being loaded into the target chamber. The Gammasphere spectrometer [37], consisting of 100 HPGe detectors, each in its own bismuth germinate (BGO) escape suppression shield, was used with a trigger accepting γγγ and higher coincidences. The sorting of the data resulted in $2.05\times10^9$ γγγ (or higher fold) coincident events formed into a standard cube that was analyzed using the *Radware* suite of programs [38].

While the data were of high statistical quality, they did suffer from over 40 contaminant reactions, both prompt and β-decay delayed. These were;

(a) The ($^{12}$C,xαyn) reactions arising from the fact that $^{12}$C is easily split into three α particles by collisions with the target.
(b) Reactions on the $^{16}$O arising from both the Nd target and the Pb backing having significant amounts of oxidation. The γ-rays from these reactions are Doppler broadened with widths of up to 100 keV and this affects gates set on higher-energy γ-rays between 1 and 2 MeV.
(c) The reactions from the 65 MeV $^{12}$C beam that was just above the Coulomb barrier for the Pb isotopes. This produced some reaction γ-rays and significant amounts of γ-rays from the β-decays of the products.

The spin and parity assignments of the levels are based on spin-parity selection rules for low-multipolarity γ-ray transitions, the Directional Correlation from Orientated states (DCO) method [39] and the mixing of levels of the same spin-parity that lie energetically close to each other. The angular intensity ratio for Gammasphere in this experiment is;

$$R_{DCO} = I_{\gamma 1}(near\ 30^o : gated\ on\ \gamma_2\ near\ 90^o) / I_{\gamma 1}(near\ 90^o : gated\ on\ \gamma_2\ near\ 30^o)$$

For stretched transitions, that is $J \rightarrow (J\text{-}\lambda)$ where λ is the multipolarity of the γ-ray, we expect $R_{DCO} \approx 1$ and $R_{DCO} \approx 0.6$ for E2 and dipole transitions respectively, when they are gated by $\gamma_2$ that is a stretched E2. If the gating $\gamma_2$ is a stretched dipole transition then we expect $R_{DCO} \approx 1$ and $R_{DCO} \approx 1.5$ for $\gamma_1$ being a dipole and E2 respectively. If the transitions are not stretched, or have finite M1/E2 mixing ratios, the situation is more complicated [40]. To determine parities, conversion coefficients or γ-ray polarization measurements are required [41]. The full set of DCO ratios are given in Table II together with γ-ray and level energies measured in this study.

The partial level scheme of $^{156}$Dy, showing the new positive-parity bands discussed in this work, is displayed in Fig. 1. The bands are labeled using the notation of Ref. [42], the new Bands 12, 17, and 20 being labeled in the order in which they were discovered. To identify levels on top of the new positive-parity bands found in this experiment, a previous data set was examined [43] that had used the $^{124}$Sn($^{36}$S,4n) $^{156}$Dy reaction, which transfers more angular momentum in the compound system than the present experiment. The current level scheme further confirms and strengthens the placements of the structures that have been proposed by previous in-beam work [44,45] from low to high-spin, and it contributes mostly to non-yrast states in the medium to high spin range.

## III. EXPERIMENTAL RESULTS

### A. The ground-state band

We observe the ground-state rotational band up to the $28^+$ state, as shown in Fig. 1. Previous experiments using the $^{124}$Sn($^{36}$S,4n)$^{156}$Dy reaction [43] have firmly established this band up to spin $58^+$. We observe three new γ-rays decaying out of this band at 458, 679 and 834 keV from the $18^+$, $20^+$ and $22^+$ levels respectively. These feed the yrast aligned S-band where these bands cross.

### B. The $\nu(i_{13/2})^2$ aligned S-band

This band was first observed by Andrews *et al.* [46] and becomes the yrast band at spin $16^+$. At spin $12^+$ it crosses the rotational band based on the 676 keV $0_2^+$ band head. New decays of 503 keV are observed from the $18^+$ level to the ground-state band and of 259 keV from the $12^+$ level to the $10^+$ state in the even γ-band (Fig. 1 and Table II). We observe the S-band band to spin $30^+$ and it has been firmly identified to spin $58^+$ [43].

### C. Gamma-band extensions

The current experiment confirms all the transitions that have been previously observed [44,45] up to spins 10 and 15 in both the even- and odd-spin γ-band respectively. In addition, the even-spin sequence of the γ-band has been extended by nine in-band transitions up to $28^+$, while its odd-spin counterpart has six new transitions to $27^+$. The $R_{DCO}$ values were measured for some of the in-band members, where possible, and they are consistent with an E2 character (Table II). Moreover, to further complement the high-spin structures of these states, the $^{36}$S induced reaction data have been reanalyzed. The data for this experiment were collected more than a decade ago by Kondev *et al.* [43]. These data have good statistics and not only enabled us to confirm the high-spin states on the γ-bands, observed in the present experiment, but also allowed us to further extend the γ-bands by two levels for both the even- and odd-spin γ sequences. Figs. 2(a) and (b), show the spectra that were used to establish new in-band members of both the even- and odd-spin sequences of the γ-bands. The inserts in both Figs. 2(a) and (b), labelled as (i) and (ii), show spectra from the $^{36}$S induced reaction for both the even- and odd-spin sequences respectively. The spin and parity assignment for the new levels to $32^+$ and $31^+$ are based on the assumption that these additional levels are connected by stretched E2 transitions. The extensions that have been made on these sequences from the current analysis mark the highest spin ever to be observed for both the even- and odd-spin γ-bands in any nucleus.

### D. New positive parity bands

#### D.1 Band 12

Band 12 is a new structure that has been established from this work using the carbon-induced reaction. A representative spectrum showing some of the in-band members of Band 12 is presented in Fig. 3(a). This rotational sequence consists of six in-band transitions. The main intensity of this band flows through high energy 1125, 1132, 1176 and 1270 keV out-of-band transitions that populate the ground-state band (Fig. 1). The $R_{DCO}$ values for these high-energy transitions were found to be close to one (Table II) and this implies that they are likely to be stretched E2 transitions or un-stretched E1 transitions. If these γ-rays are assigned as un-stretched E1 transitions it would mean that the other out-of-band 451, 520 and 636 keV γ-rays are M2 transitions, and this is considered very unlikely. Therefore, we conclude that these high-energy transitions are stretched E2 transitions. This gives spin-parity assignments of $16^+$, $18^+$, $20^+$, $22^+$, $24^+$, $26^+$ and $28^+$ to the 4157, 4697, 5309, 5983, 6721, 7519 and 8369 keV levels of Band 12 respectively. Furthermore, the energies of the $16^+$ states of this band and the even γ-band are close in energy (i.e., 4157 and 4110 keV respectively).

The interlinking 586 and 573 keV E2 transitions that connect these two bands indicate that there is a mixing between these bands at this spin 16. This is only possible if these bands are of the same spin-parity and this confirms our spin and parity assignments.

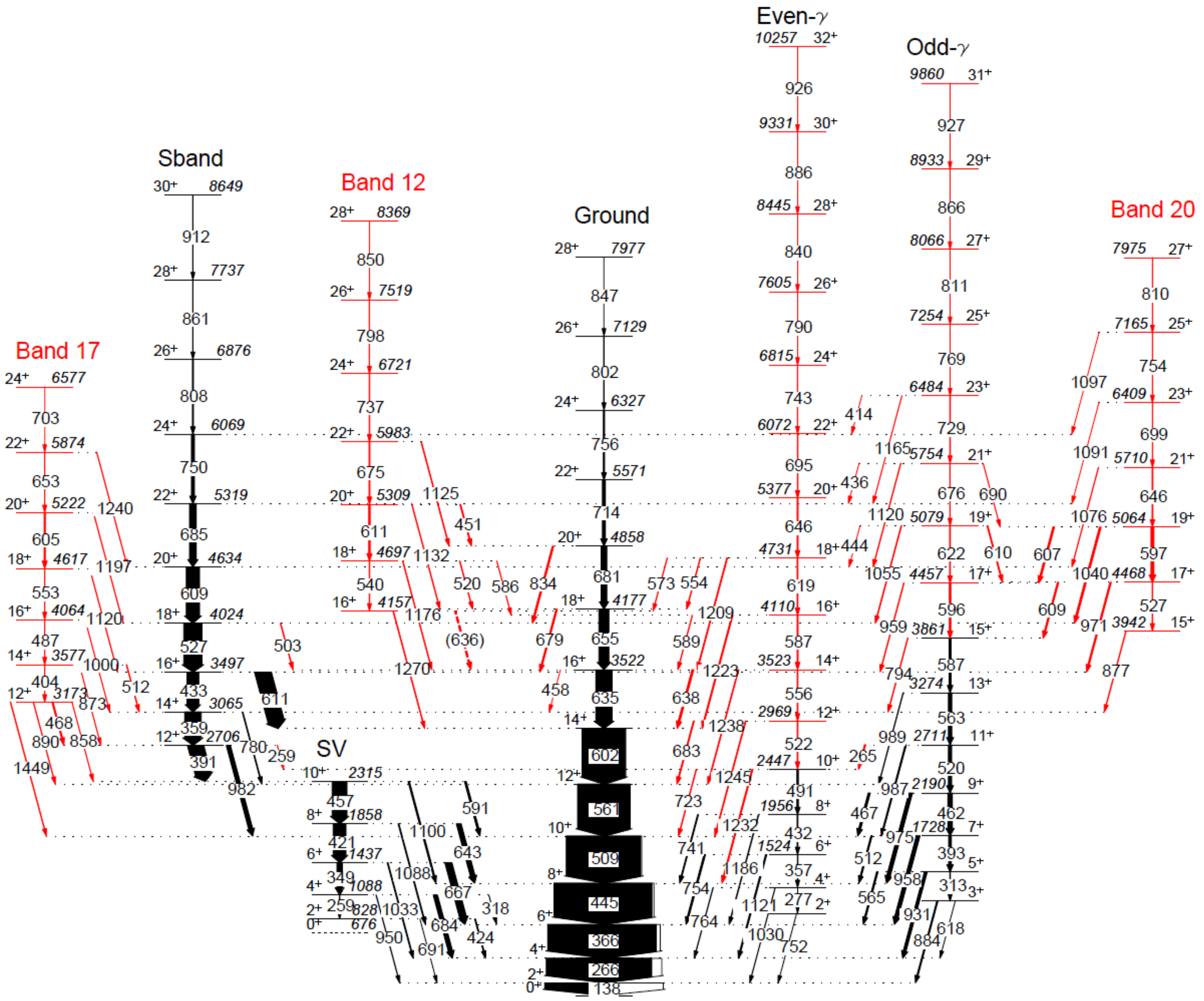

*FIG. 1. (Color on-line) Partial level scheme deduced for $^{156}$Dy from the current work for positive-parity states. New levels and γ-ray transitions are shown in red while their known counterparts, from previous in-beam work, are in black. The widths of the arrows are proportional to the intensities of the transitions.*

### D.2. Band 17

Band 17 is a new band that decays mainly to the S-band (Fig. 1). The DCO ratios of the 873 and 1000 keV transitions feeding the 12+ and 14+ members of the S-band show stretched E2 character (Table II) highly favoring the 3577 and 4064 keV members of band 17 to be 14+ and 16+, respectively. Again, if these transitions were to be assigned as unstretched E1 transitions it would mean that the other out-of-band 512, 468 and 890 keV transitions are M2 transitions, and this is very unlikely. The assignments of the rest of the levels are based on the assumption that they are band members of Band 17 connected by in-band E2 transitions. A coincidence spectrum illustrating the photo peaks corresponding to the in-band members of this structure is shown in Fig. 3(b).

### D.3. Band 20

A previously unobserved sequence of seven levels based upon the state at 3942 keV has been established and is labeled as Band 20 in this work. This band is also amongst the newly identified medium-spin structures that decay predominantly to the S-band via a series of high-energy transitions. It also decays to the $17^+$ and $15^+$ members of the odd-spin γ-band via almost degenerate 607 and 609 keV transitions, respectively. The DCO measurements for the majority of the in-band members of this structure are consistent with them being stretched E2 transitions. At spin $17^+$, cross transitions link the odd-spin γ-band and Band 20 from almost energy degenerate states (i.e., separated by less than 20 keV). The identification of the interconnecting transitions between sequences of the same parity and signature confirm and strengthen the proposed assignments. A double-gated spectrum showing some of the in-band members of this structure is shown in Fig. 3(c).

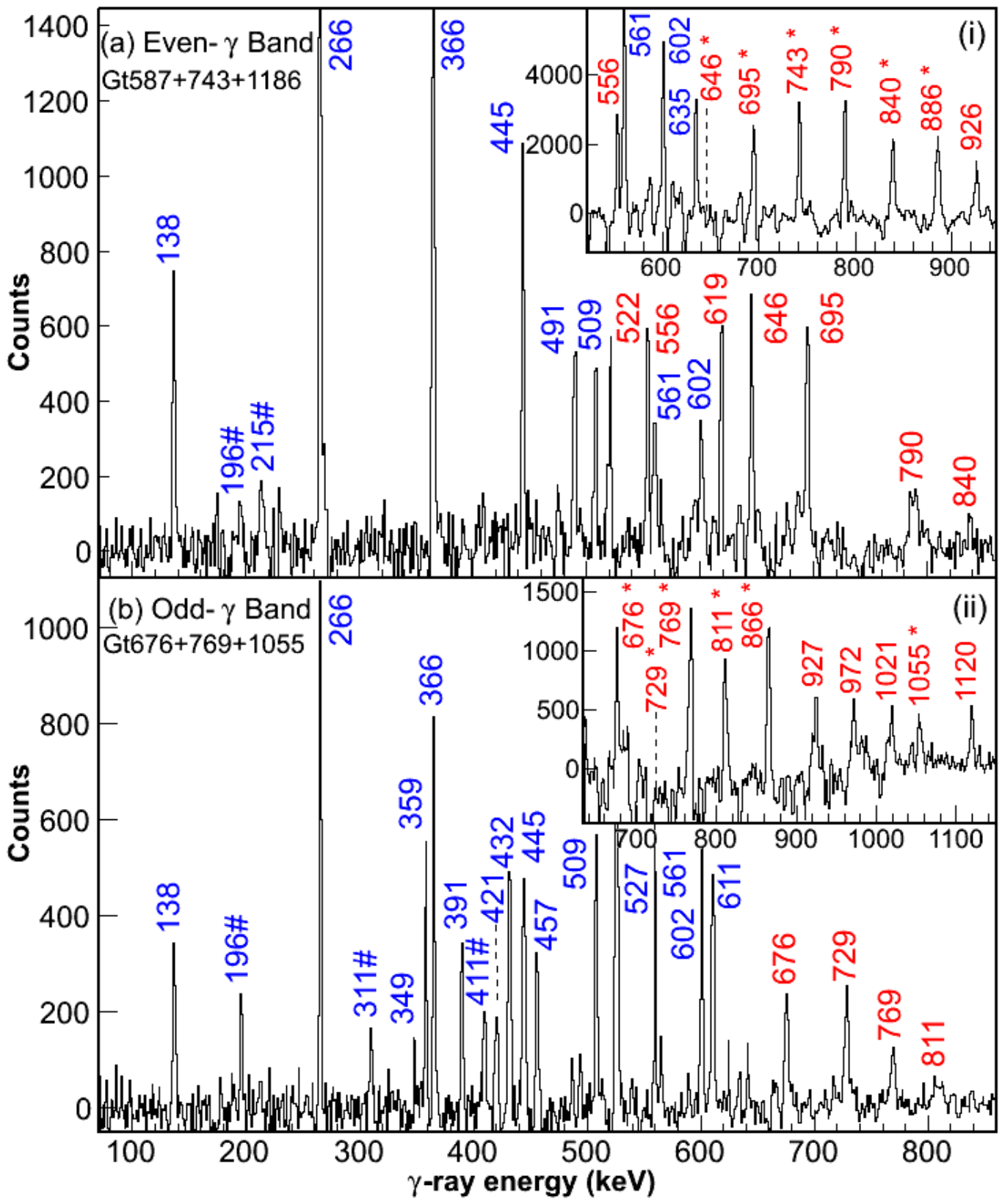

*FIG. 2. (Color online) Summed coincident spectra for the even-spin (a) and odd-spin (b) members of the γ-bands obtained from the $^{12}C$ and $^{36}S$ (inserts (i) and (ii)) induced reactions. New transitions are colored in red while contaminants are denoted by hash (#) symbols. In the inserts (i) and (ii), transitions marked by asterisks (*) were included in the gating that produced the coincident spectra.*

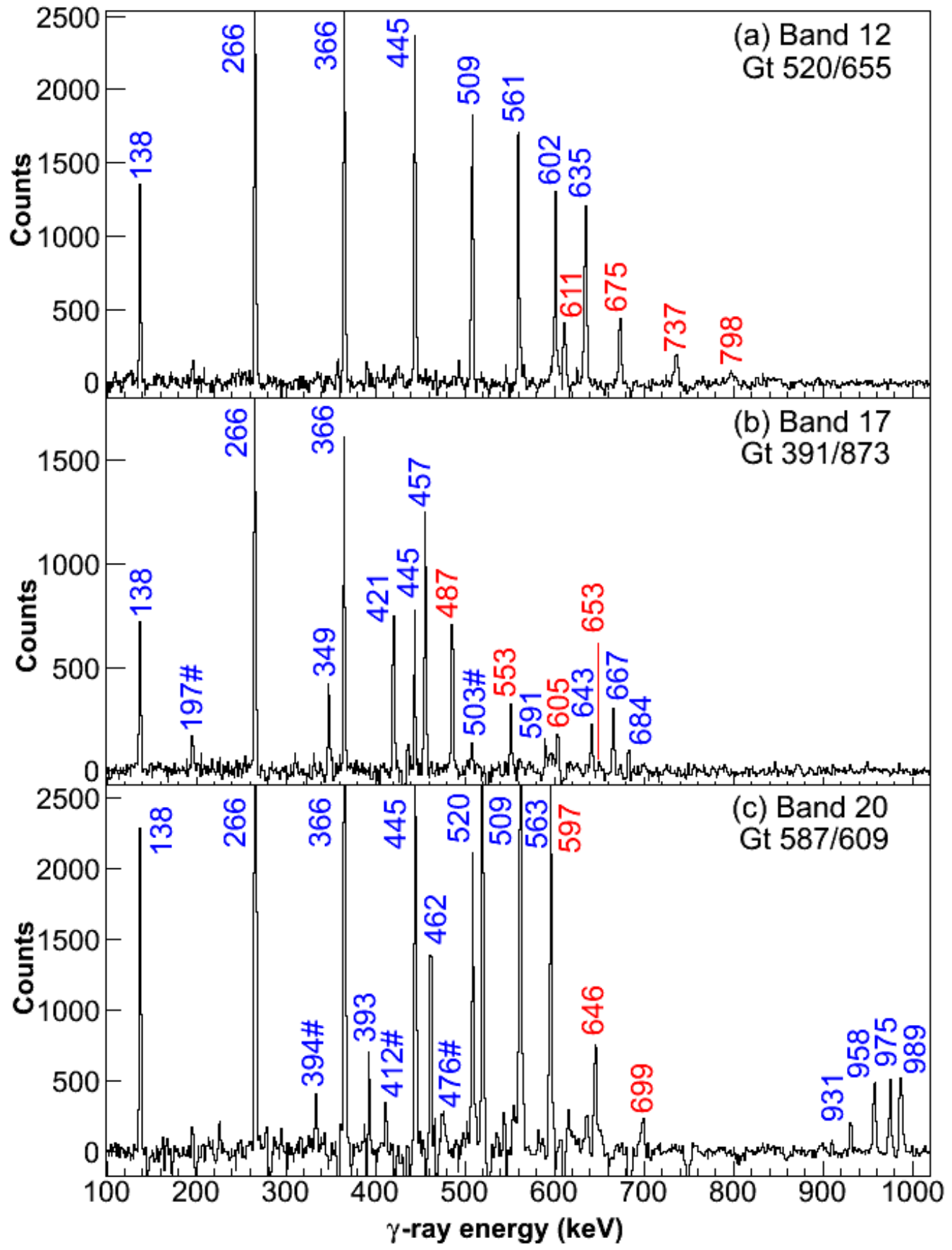

*FIG. 3. (Color online) (a), (b) and (c) Summed coincident spectra for Bands 12, 17 and 20, respectively, obtained from the $^{12}C$ data. Transitions corresponding to these newly-found structures are labeled in red while contaminants are denoted by hash (#) symbols.*

***Table II.*** *Gamma-ray $E_\gamma$ (keV) and level energies $E_x$ ( keV) of bands shown in Fig. 1 together with angular intensity ratios $R_{DCO}$ for some of the transitions. Blank entries indicate information that could not be obtained.*

| $E_x$(keV) | $E_\gamma$(keV)§ | Multipolarty | $I_i^\pi$ | $I_f^\pi$ | Band$_f$ | $R_{DCO}$ |
|---|---|---|---|---|---|---|
| **Ground** | | | | | | |
| 138 | 137.6 | E2 | $2^+$ | $0^+$ | Ground | 0.95(1) |
| 404 | 266.2 | E2 | $4^+$ | $2^+$ | Ground | 0.991(3) |
| 770 | 366.1 | E2 | $6^+$ | $4^+$ | Ground | 1.000(1) |
| 1215 | 445.1 | E2 | $8^+$ | $6^+$ | Ground | |
| 1724 | 508.9 | E2 | $10^+$ | $8^+$ | Ground | |
| 2284 | 560.7 | E2 | $12^+$ | $10^+$ | Ground | |
| 2886 | 601.8 | E2 | $14^+$ | $12^+$ | Ground | |
| 3522 | 458.2 | E2 | $16^+$ | $14^+$ | S-band | |
| | 635.4 | E2 | $16^+$ | $14^+$ | Ground | |
| 4177 | 655.1 | E2 | $18^+$ | $16^+$ | Ground | |
| | 679.2 | E2 | $18^+$ | $16^+$ | S-band | |
| 4858 | 680.6 | E2 | $20^+$ | $18^+$ | Ground | |
| | 834.2 | E2 | $20^+$ | $18^+$ | S-band | |
| 5571 | 713.6 | E2 | $22^+$ | $20^+$ | Ground | |
| 6327 | 756.1 | E2 | $24^+$ | $22^+$ | Ground | |
| 7129 | 802.1 | E2 | $26^+$ | $24^+$ | Ground | |
| 7977 | 847.3 | E2 | $28^+$ | $26^+$ | Ground | |

| $E_x$(keV) | $E_\gamma$(keV)§ | Multipolarty | $I_i^\pi$ | $I_f^\pi$ | Band$_f$ | $R_{DCO}$ |
|---|---|---|---|---|---|---|
| | | | ***Band12*** | | | |
| 4157 | 636.4 | M1/E2 | $16^+$ | $16^+$ | Ground | |
| | 1270.3 | E2 | $16^+$ | $14^+$ | Ground | 0.89(7) |
| 4697 | 520.2 | M1/E2 | $18^+$ | $18^+$ | Ground | |
| | 540.4 | E2 | $18^+$ | $16^+$ | Band 12 | |
| | 586.4 | E2 | $18^+$ | $16^+$ | Even-γ | |
| | 1175.8 | E2 | $18^+$ | $16^+$ | Ground | 1.09(11) |
| 5309 | 451.1 | M1/E2 | $20^+$ | $20^+$ | Ground | |
| | 611.1 | E2 | $20^+$ | $18^+$ | Band 12 | |
| | 1132.3 | E2 | $20^+$ | $18^+$ | Ground | 0.89(7) |
| 5983 | 674.5 | E2 | $22^+$ | $20^+$ | Band 12 | |
| | 1125.5 | E2 | $22^+$ | $20^+$ | Ground | |
| 6721 | 737.2 | E2 | $24^+$ | $22^+$ | Band 12 | |
| 7519 | 798.2 | E2 | $26^+$ | $24^+$ | Band 12 | |
| 8369 | 849.7 | E2 | $28^+$ | $26^+$ | Band 12 | |
| | | | ***Second Vacuum (SV)*** | | | |
| 828 | 424.1 | E2 | $2^+$ | $4^+$ | Ground | 0.98(2) |
| | 690.7 | M1/E2 | $2^+$ | $2^+$ | Ground | |
| 1088 | 259.5 | E2 | $4^+$ | $2^+$ | SV | |
| | 317.7 | E2 | $4^+$ | $6^+$ | Ground | |
| | 683.8 | M1/E2 | $4^+$ | $4^+$ | Ground | |
| | 950.4 | E2 | $4^+$ | $2^+$ | Ground | |
| 1437 | 348.8 | E2 | $6^+$ | $4^+$ | SV | 0.981(3) |
| | 666.7 | M1/E2 | $6^+$ | $6^+$ | Ground | 0.682(3) |
| | 1032.9 | E2 | $6^+$ | $4^+$ | Ground | |
| 1858 | 421.3 | E2 | $8^+$ | $6^+$ | SV | |
| | 642.9 | M1/E2 | $8^+$ | $8^+$ | Ground | 0.712(3) |
| | 1088.1 | E2 | $8^+$ | $6^+$ | Ground | |
| 2315 | 456.6 | E2 | $10^+$ | $8^+$ | SV | |
| | 590.7 | M1/E2 | $10^+$ | $10^+$ | Ground | 0.71(1) |
| | 1099.7 | E2 | $10^+$ | $8^+$ | Ground | |
| | | | | | | |
| | | | ***S-band*** | | | |
| 2706 | 258.8 | E2 | $12^+$ | $10^+$ | Odd-γ | |
| | 390.9 | E2 | $12^+$ | $10^+$ | SV | |
| | 981.7 | E2 | $12^+$ | $10^+$ | Ground | |
| 3065 | 359.1 | E2 | $14^+$ | $12^+$ | S-band | 1.000(3) |
| | 780.3 | E2 | $14^+$ | $12^+$ | Ground | |
| 3497 | 432.7 | E2 | $16^+$ | $14^+$ | S-band | |
| | 611.2 | E2 | $16^+$ | $14^+$ | Ground | |
| | 502.5 | E2 | $18^+$ | $16^+$ | S-band | |
| | 526.9 | E2 | $18^+$ | $16^+$ | S-band | |
| 4634 | 609.5 | E2 | $20^+$ | $18^+$ | S-band | |
| 5319 | 684.7 | E2 | $22^+$ | $20^+$ | S-band | |
| 6069 | 750.1 | E2 | $24^+$ | $22^+$ | S-band | |
| 6876 | 807.6 | E2 | $26^+$ | $24^+$ | S-band | |
| 7737 | 861.2 | E2 | $28^+$ | $26^+$ | S-band | |
| 8649 | 912.1 | E2 | $30^+$ | $28^+$ | S-band | |
| | | | ***Odd-spin γ-band ( Odd-γ)*** | | | |
| 1022 | 617.7 | M1/E2 | $3^+$ | $4^+$ | Ground | |
| | 884.3 | M1/E2 | $3^+$ | $2^+$ | Ground | |
| 1335 | 312.9 | E2 | $5^+$ | $3^+$ | Odd-γ | |
| | 565.4 | M1/E2 | $5^+$ | $6^+$ | Ground | |
| | 931.1 | M1/E2 | $5^+$ | $4^+$ | Ground | |
| 1728 | 393.1 | E2 | $7^+$ | $5^+$ | Odd-γ | |
| | 512.4 | M1/E2 | $7^+$ | $8^+$ | Ground | |
| | 958.1 | M1/E2 | $7^+$ | $6^+$ | Ground | |
| 2190 | 462.5 | E2 | $9^+$ | $7^+$ | Odd-γ | |

| $E_x(keV)$ | $E_\gamma(keV)$[§] | Multipolarty | $I_i^\pi$ | $I_f^\pi$ | $Band_f$ | $R_{DCO}$ |
|---|---|---|---|---|---|---|
| **Odd-spin γ-band ( Odd-γ)** | | | | | | |
| | 466.5 | M1/E2 | $9^+$ | $10^+$ | Ground | |
| | 975.4 | M1/E2 | $9^+$ | $8^+$ | Ground | |
| | 520.1 | E2 | $11^+$ | $9^+$ | Odd-γ | |
| | 986.6 | M1/E2 | $11^+$ | $10^+$ | Ground | |
| 3274 | 563.1 | E2 | $13^+$ | $11^+$ | Odd-γ | |
| 3274 | 989.2 | M1/E2 | $13^+$ | $12^+$ | Ground | |
| 3861 | 587.2 | E2 | $15^+$ | $13^+$ | Odd-γ | |
| | 794.4 | M1/E2 | $15^+$ | $14^+$ | SV | |
| 4457 | 595.7 | E2 | $17^+$ | $15^+$ | Odd-γ | 0.902(12) |
| 4457 | 959.3 | M1/E2 | $17^+$ | $16^+$ | SV | |
| 5079 | 443.7 | M1/E2 | $19^+$ | $18^+$ | Even-γ | |
| | 610.3 | M1/E2 | $19^+$ | $17^+$ | Band 20 | |
| | 621.8 | E2 | $20^+$ | $18^+$ | Odd-γ | |
| | 1055.3 | M1/E2 | $20^+$ | $18^+$ | S-band | |
| 5754 | 436.1 | M1/E2 | $21^+$ | $20^+$ | Odd-γ | |
| | 676.5 | E2 | $21^+$ | $19^+$ | Odd-γ | |
| | 690.1 | E2 | $21^+$ | $19^+$ | Band 20 | |
| | 1120.3 | M1/E2 | $21^+$ | $20^+$ | S-band | |
| 6484 | 414.1 | M1/E2 | $23^+$ | $22^+$ | Odd-γ | |
| | 729.5 | E2 | $23^+$ | $21^+$ | Odd-γ | |
| | 1164.9 | M1/E2 | $23^+$ | $22^+$ | S-band | |
| 7254 | 769.5 | E2 | $25^+$ | $23^+$ | Odd-γ | |
| 8066 | 811.2 | E2 | $27^+$ | $25^+$ | Odd-γ | |
| 8933 | 865.5 | E2 | $29^+$ | $27^+$ | Odd-γ | |
| 9860 | 926.9 | E2 | $31^+$ | $29^+$ | Odd-γ | |
| **Even-spin γ-band ( Even-γ)** | | | | | | |
| 890 | 752.4 | M1/E2 | $2^+$ | $2^+$ | Ground | |
| 1168 | 277.2 | E2 | $4^+$ | $2^+$ | Even-γ | |
| | 763.9 | M1/E2 | $4^+$ | $4^+$ | Ground | |
| | 1030.4 | E2 | $4^+$ | $2^+$ | Ground | |
| 1524 | 356.5 | E2 | $6^+$ | $4^+$ | Even-γ | |
| | 754.5 | M1/E2 | $6^+$ | $6^+$ | Ground | |
| | 1120.6 | E2 | $6^+$ | $4^+$ | Ground | |
| 1956 | 431.9 | E2 | $8^+$ | $6^+$ | Even-γ | |
| | 741.1 | M1/E2 | $8^+$ | $8^+$ | Ground | |
| | 1186.3 | E2 | $8^+$ | $6^+$ | Ground | |
| 2447 | 490.5 | E2 | $10^+$ | $8^+$ | Even-γ | |
| | 722.7 | M1/E2 | $10^+$ | $10^+$ | Ground | |
| | 1231.5 | E2 | $10^+$ | $8^+$ | Ground | |
| 2969 | 522.1 | E2 | $12^+$ | $10^+$ | Even-γ | |
| | 682.9 | M1/E2 | $12^+$ | $12^+$ | Ground | |
| | 1245.1 | E2 | $12^+$ | $10^+$ | Ground | |
| 3523 | 555.7 | E2 | $14^+$ | $12^+$ | Even-γ | |
| | 638.1 | M1/E2 | $14^+$ | $14^+$ | Ground | |
| | 1238.2 | E2 | $14^+$ | $12^+$ | Ground | |
| 4110 | 587.3 | E2 | $16^+$ | $14^+$ | Even-γ | 0.99(1) |
| | 589.2 | E2 | $16^+$ | $16^+$ | Ground | |
| | 1223.1 | E2 | $16^+$ | $14^+$ | Ground | |
| 4731 | 553.9 | M1/E2 | $18^+$ | $18^+$ | Ground | |
| | 573.3 | E2 | $18^+$ | $16^+$ | Band 12 | |
| | 619.4 | E2 | $18^+$ | $16^+$ | Even-γ | |
| | 1209.1 | E2 | $18^+$ | $16^+$ | Ground | |
| 5377 | 646.1 | E2 | $20^+$ | $18^+$ | Even-γ | |
| 6072 | 695.5 | E2 | $22^+$ | $20^+$ | Ground | |
| 6815 | 743.2 | E2 | $24^+$ | $22^+$ | Even-γ | |
| 7605 | 790.3 | E2 | $26^+$ | $24^+$ | Even-γ | |
| 8445 | 839.9 | E2 | $28^+$ | $26^+$ | Even-γ | |

| *$E_x$(keV)* | *$E_\gamma$(keV)*[§] | *Multipolarty* | $I_i^\pi$ | $I_f^\pi$ | *$Band_f$* | *$R_{DCO}$* |
|---|---|---|---|---|---|---|
| | | | ***Even-spin γ-band ( Odd-γ)*** | | | |
| 9331 | 886.2 | E2 | $30^+$ | $28^+$ | Even-γ | |
| 10257 | 926.12 | E2 | $32^+$ | $30^+$ | Even-γ | |
| | | | ***Band 17*** | | | |
| 3173 | 467.6 | M1/E2 | $12^+$ | $12^+$ | S-band | |
| | 858.2 | E2 | $12^+$ | $10^+$ | SV | |
| | 889.7 | M1/E2 | $12^+$ | $12^+$ | Ground | |
| | 1449.1 | E2 | $12^+$ | $10^+$ | Ground | |
| 3577 | 404.5 | E2 | $14^+$ | $12^+$ | Band 17 | |
| | 511.7 | M1/E2 | $14^+$ | $14^+$ | S-band | 0.72(2) |
| | 872.5 | E2 | $14^+$ | $12^+$ | S-band | 0.94(9) |
| 4064 | 487.1 | E2 | $16^+$ | $14^+$ | Band 17 | |
| | 1000.1 | E2 | $16^+$ | $14^+$ | S-band | 0.95(6) |
| 4617 | 552.6 | E2 | $18^+$ | $16^+$ | Band 17 | |
| | 1120.1 | E2 | $18^+$ | $16^+$ | S-band | |
| 5222 | 604.5 | E2 | $20^+$ | $18^+$ | Band 17 | |
| | 1197.1 | E2 | $20^+$ | $18^+$ | S-band | |
| 5874 | 652.8 | E2 | $22^+$ | $20^+$ | Band 17 | |
| | 1240.2 | E2 | $22^+$ | $20^+$ | S-band | |
| 6577 | 703.1 | E2 | $24^+$ | $22^+$ | Band 17 | |
| | | | ***Band 20*** | | | |
| 3942 | 877.3 | M1/E2 | $15^+$ | $14^+$ | S-band | |
| 4468 | 526.9 | E2 | $17^+$ | $15^+$ | Band 20 | |
| | 609.5 | M1/E2 | $17^+$ | $16^+$ | S-band | |
| | 970.7 | E2 | $17^+$ | $15^+$ | Odd-γ | |
| 5064 | 597.4 | E2 | $19^+$ | $17^+$ | Band 20 | 0.97(2) |
| | 606.8 | E2 | $19^+$ | $17^+$ | Odd-γ | |
| | 1039.7 | M1/E2 | $19^+$ | $18^+$ | S-band | |
| 5710 | 645.9 | E2 | $21^+$ | $19^+$ | Band 20 | 1.01(1) |
| 5710 | 1076.1 | M1/E2 | $21^+$ | $20^+$ | S-band | |
| 6409 | 698.7 | E2 | $23^+$ | $21^+$ | Band 20 | 0.95(6) |
| | 1091.1 | M1/E2 | $23^+$ | $22^+$ | S-band | |
| 7165 | 753.8 | E2 | $25^+$ | $23^+$ | Band 20 | |
| | 1097.3 | M1/E2 | $25^+$ | $24^+$ | S-band | |
| 7975 | 810.5 | E2 | $27^+$ | $25^+$ | Band 20 | |
| | | | | | | |

[§]The $\gamma$ -ray energies are estimated to be accurate to ±0.3 keV.

## III. DISCUSSION

### A. The $0_2^+$ pairing isomer state at 676 keV

Recent work has shown that the first excited $0_2^+$ states in N = 88 and 90 nuclei are not β-vibrations [47], but have properties resembling pairing isomers [48]; states lowered into the pairing gap by configuration-dependent pairing [14,49]. They are neutron 2p-2h seniority zero states formed by raising two neutrons out of the ground-state configuration into the $\nu h_{11/2}$ [505]11/2$^-$ Nilsson orbit. This orbit is raised from the lower filled shell to the Fermi surface at the onset of deformation after the N = 82 shell closure. Pairs of neutrons in this high-K "oblate" polar orbit do not take part in the normal monopole pairing produced by the high density of low-Ω "prolate" equatorial rotationally aligned orbitals [58]. Due to the small overlap of the wave-functions of prolate and oblate single particle states, $[505]^2$ configurations are decoupled from the pairing involved in the alignment of low-Ω $i_{13/2}$ neutrons that cause back-bending in the rotational bands of deformed nuclei near N = 90 [50]. The first excited $0_2^+$ states in N = 88 and 90 nuclei are then seniority zero states. These $0_2^+$ states may be viewed as a "second vacuum" [14,49] on which excited states are built that are congruent to the excited states built on the ground-state vacuum. We will refer to the rotational band built on the $0_2^+$ state at 676 keV in $^{156}$Dy as the second vacuum band (SV).
As observed by Andrews *et al.* [46], this SV band is crossed by the S-band at spin 12$^+$ and at a rotational frequency of ~200 keV compared to the usual $i_{13/2}$ neutron alignment frequency $\hbar\omega_c \approx 280$ keV [51]. This crossing is just a simple band crossing and has nothing to do with the Cranked Shell Model [52] *AB* band crossings [53] where the Coriolis force exceeds the monopole pairing energy.

### B. The $K^\pi = 2^+$ γ-bands

In Fig. 4 we show the excitation energies $E_x$ minus a rotational energy $E_R = 7.7I(I+1)$ MeV for the positive-parity states that we observe (Fig. 1) in $^{156}$Dy. In Fig. 5(a) we show the alignment $i_x = I_x(\omega)-I_{x,ref}(\omega)$ of these bands plotted against rotational frequency $\hbar\omega$. In Fig. 5(b) we show the energies, $e'$ Routhians [52], in the rotating frame for these bands, again as a function of $\hbar\omega$.

It can be seen from Fig. 4 that both the even- and the odd-spin γ-bands track parallel to the ground-state band faithfully all the way to spin 32$^+$. This is very similar to the observation of Ollier *et al.* [9] who find the odd-spin γ-band in $^{160}$Er to track parallel to the yrast band through both the first backbend, $i_{13/2}$ neutron alignment, and then through the second up-bend, $h_{11/2}$ proton alignment. In Fig. 5(a), where the alignment $i_x$ is plotted against rotational frequency $\hbar\omega_c$, it can be seen that the alignment in the γ-bands comes at a slightly lower frequency than in the ground-state band. This has been explained by Yates *et al.* [11] as a geometrical effect due to the additional excitation energies of the γ-bands and the aligned $i_{13/2}$ neutron bands based on the γ-bands.

The even-spin members of the γ-band in $^{156}$Dy have more neighbouring bands of the same spin parity than the odd-spin members of the γ-band. Hence we would expect there to be more band mixing affecting the excitation energies of the positive-parity states than odd-spin positive-parity bands disturbing the energies of the odd-spin states. It can be seen from Fig. 4 that there is no signature splitting in the γ-bands up to spin 12$^+$, after which there is a very small signature splitting where the even-spin states lie below the odd-spin states. This could be due to either different band mixing or due to a change in the γ deformation. The absence of signature splitting is an indication that there is no γ-deformation so that the nucleus is axially symmetric.

## C. Bands 17 and 20

Both Band 17 and Band 20 decay mainly to the yrast S-band. The exceptions are two transitions to the ground-state band out of the lowest, $12^+$ state of Band 17 and where Band 20 crosses the odd-spin γ-band between spins $17^+$ and $21^+$. This strongly preferred decay pattern suggests that these two bands are closely connected to the S-band. Indeed, Fig. 5(a) shows that both Band 17 and Band 20 have alignments close to 10ħ. We suggest that Band 17 and Band 20 could be bands formed by a γ vibration built on the S-band configuration. In Fig. 6 we compare the ratios of the out-of-band *B(E2;out)* to in-band *B(E2;in)* strengths

$$R_{E2} = B\,(E2;out)/\,B(E2;in)$$

for the γ decays from the γ-bands to the ground-state band with the γ decays from Bands 17 and 20 to the S-band. We assume that the out-of-band $J \rightarrow (J\text{-}1)$ transitions are dominated by the E2 component due to the $\Delta K = 2$ nature of γ-band to ground-state band transitions [2,54]. It can be seen that these ratios are of the same order of magnitude giving support to the interpretation of the structure of Bands 17 and 20. Band 20 crosses the odd-spin γ-band between spins $17^+$ and $19^+$ which will cause some degree of mixing and affect branching ratios and transitions rates. Similarly, there is a mixing of even-spin structures near spin $16^+$ that will influence the transition rates from Band 17. There is a small signature splitting between Bands 17 and 20 with the even-spin states lying lowest in energy. Again, this could either indicate that the S-band configuration is somewhat γ deformed, or that these even-spin states mix with nearby states of the same spin and parity. It should be noted that after the alignment of two $i_{13/2}$ neutrons their core has become $^{154}$Dy. The γ-band in $^{154}$Dy shows significant signature splitting [55], also with the even-spin states lying lowest in energy, indicating that it is γ deformed. This shape of the core for the $^{156}$Dy S-band could be a reason for Bands 17 and 20 showing signature splitting.

## D. Band 12

Band 12 has no observable signature partner and decays only to the ground-state band with no observed decay branches to other structures. Fig. 5(a) shows that Band 12 is consistent with a configuration having two aligned $i_{13/2}$ neutrons. An even-spin structure at this excitation energy that has no signature partner could be the S-band built on the $0_2^+$ second vacuum configuration. This would be the first observation of such a structure. Band 12 has both the correct excitation energy and the correct alignment to support this interpretation. If this is the case, then the plot of the Routhians $\boldsymbol{e'}$ shown in Fig. 5(b) can be used to estimate the crossing frequency [50] $\boldsymbol{\hbar\omega_c}$ of the two bands. The projection of the SV band is found to cross band 12 at $\boldsymbol{\hbar\omega_c}$ = 0.28(1) MeV. The average unattenuated "AB" $i_{13/2}^{\,2}$ aligned S-band band crossing in this region of nuclei is $\boldsymbol{\hbar\omega_c}$ = 0.27(1) MeV [51]. This would suggest that this postulated band crossing is not blocked by the nucleons making up the configuration of the second vacuum. As the configuration of the SV band has large components based on the $\upsilon\{[505]11/2^-\}^2$ configuration and the $[505]11/2^-$ neutrons do not partake in the normal monopole pairing [50], the normal band crossing frequency will not be blocked [51] in agreement with our conjecture.

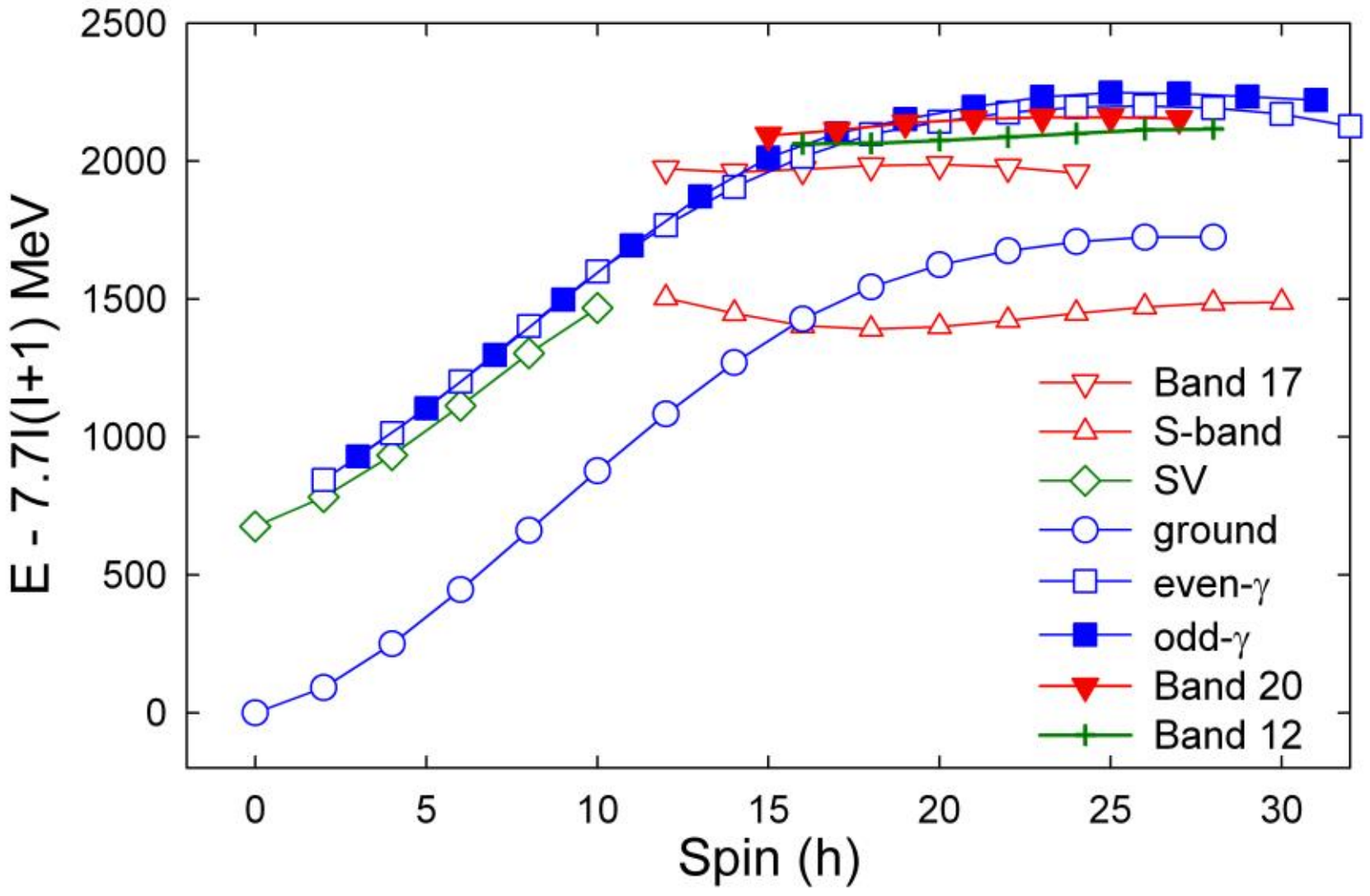

*Fig. 4. (Color online) Plot of the excitation energy minus a rigid rotor for the positive-parity bands shown in Fig. 1.*

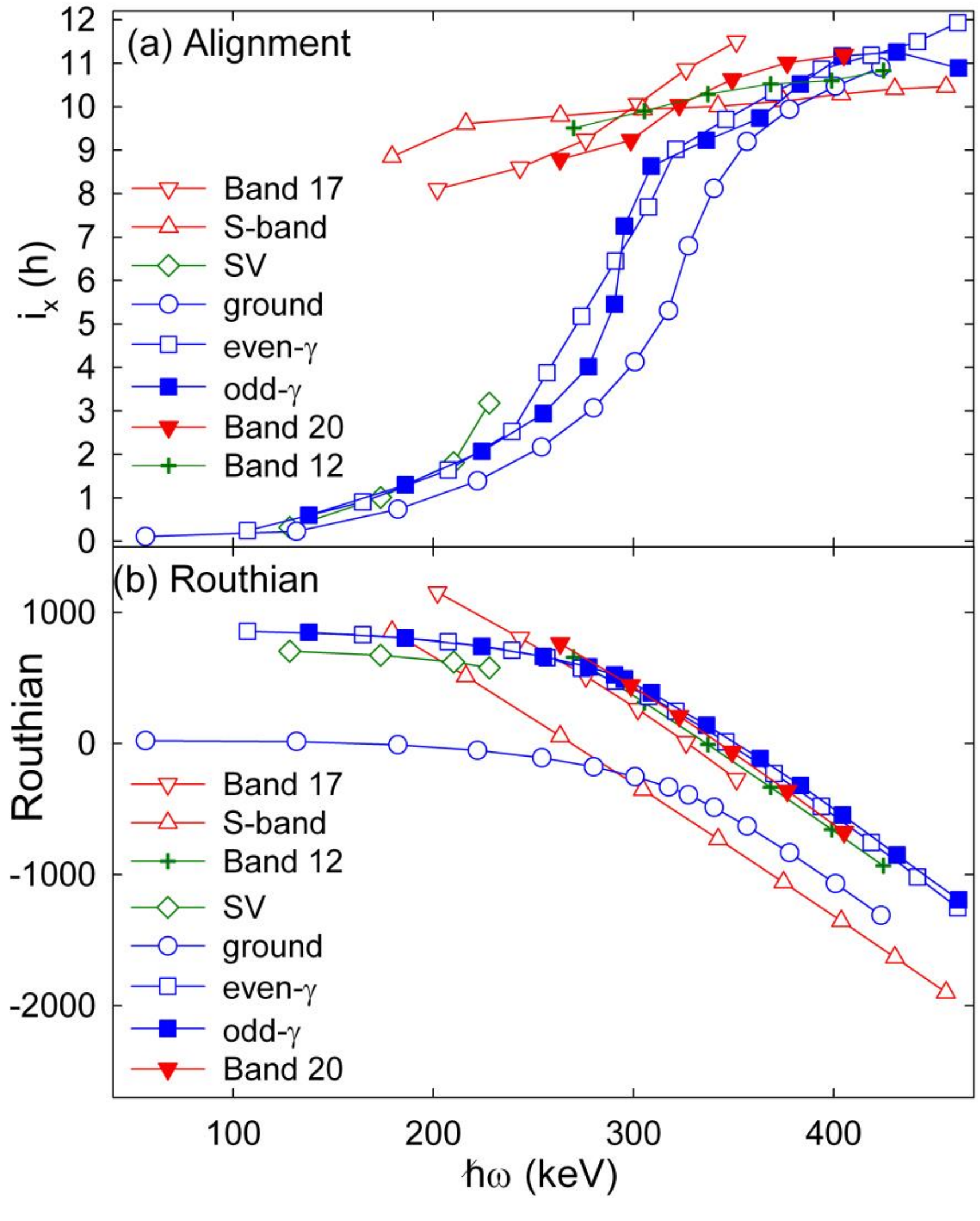

*Fig. 5. (Color online) (a) and (b) Plots of alignment $i_x$ and Routhians e' respectively for the positive-parity bands in $^{156}$Dy.*

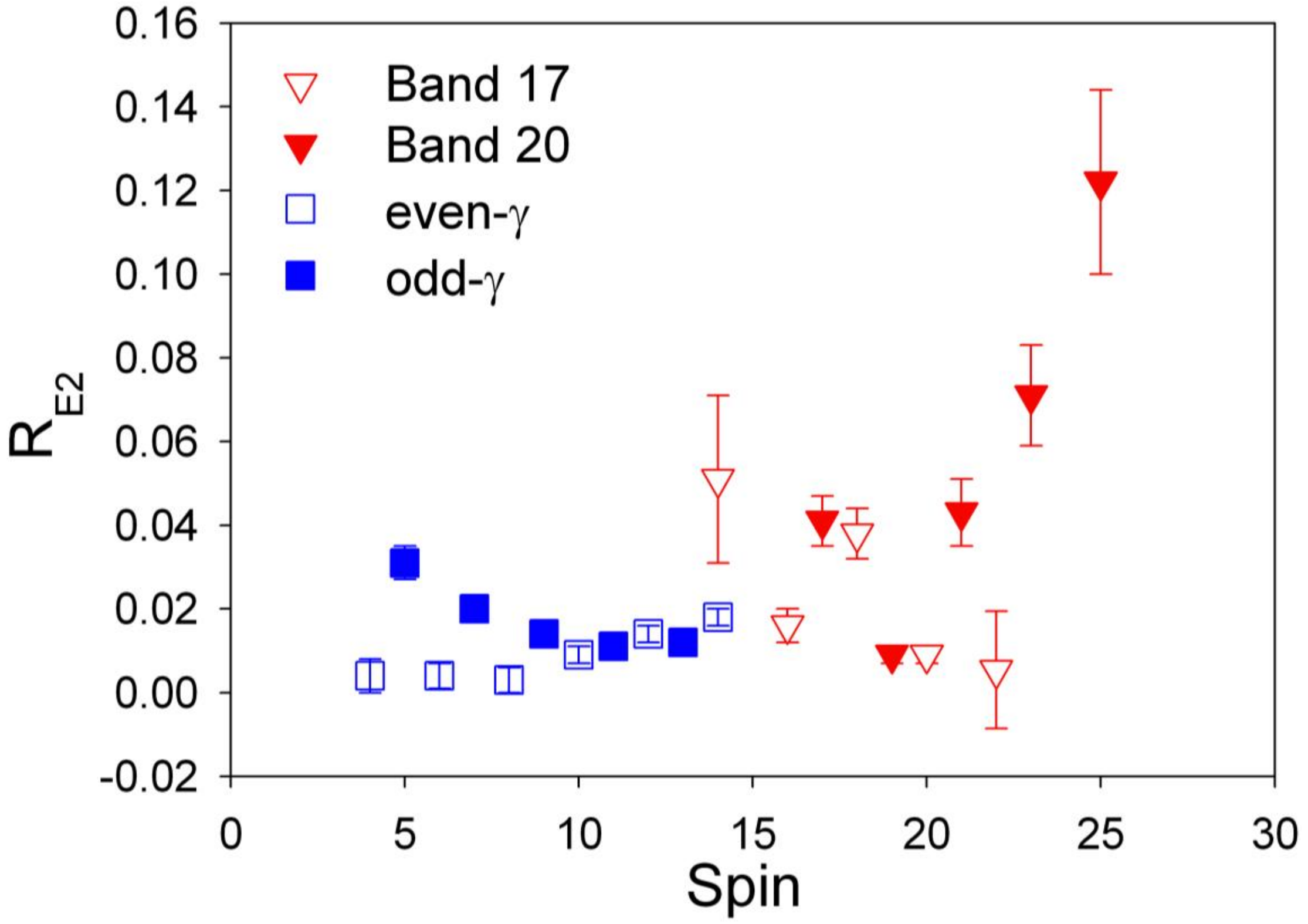

*Fig. 6. (Color online) Ratios of the B(E2) values for out-of-band to in-band transitions, $R_{E2}$ = B(E2;out)/ B(E2;in), for the γ decays from the γ-bands to the ground-state band with the γ decays from Bands 17 and 20 to the S-band. It is assumed that the out-of-band J → (J-1) transitions are dominated by the E2 component due to the ΔK = 2 nature of γ-band to ground-state band transitions, so we have neglected any M1 components.*

## IV CONCLUSIONS

We have investigated the positive-parity structures in $^{156}$Dy at medium spins using the $^{148}$Nd ($^{12}$C, 4n) reaction at 65 MeV with the Gammasphere spectrometer at the Argonne National Laboratory. Significant extensions have been made for the γ-band to spins $32^+$ and $31^+$. The new levels that have been established for both even- and odd-spin γ-band sequences, from the current analysis, mark the highest spin ever to be observed for both signatures a γ-band in any nucleus. These bands are also found to faithfully track the ground-state configuration, which is a general feature of γ-bands in the transitional rare-earth region.

Three new bands have been discovered, all with an alignment of ~10ħ. It is proposed that the even-spin Band 17 and the odd-spin Band 20 are the even- and odd-spin members of a γ-band built on the $i_{13/2}{}^2$ aligned S-band. This is the first example of the observation of γ vibrations built on both the continuation of the ground-state configuration and on the aligned S-band.

The other newly observed band, Band 12, is conjectured to be the $i_{13/2}{}^2$ aligned $S_2$-band built on the $0_2^+$ second vacuum configuration. The crossing of this band with the band built on the $0_2^+$ state is at a rotational frequency $\hbar\omega_c$ = 0.28(1) MeV which is consistent with the second vacuum configuration not blocking the monopole pairing.

The new experimental results on γ-bands presented here are of particular interest because of new interpretations of $K^\pi=2^+$ rotational bands in even-even nuclei. In the symplectic shell model [56] $K^\pi=2^+$ rotational bands arise naturally and are not due to vibrations in the γ degree of freedom. Recent studies [57] with the algebraic version of the Bohr model, for which calculation without invoking the adiabatic approximation are now feasible, lend strong support for the view [14,49] that low-energy rotational bands are close to being β rigid. It also indicates that they should be near γ rigid as well and that the low-energy $K^\pi=2^+$ rotational bands are more meaningfully described as rotations with small triaxial deformation of a strongly-renormalized SU(3) type.

## ACKNOWLEDGEMENTS

We would like to thank the crew of the ANL ATLAS accelerator for delivering a very stable and clean beam. The authors also thank the ANL operations staff at Gammasphere and gratefully acknowledge the efforts of J.P. Green for target preparation. In addition we thank David Radford for software support and Sven Åberg for constructive discussions. This work is funded by the U.S. National Science Foundation under Grant Nos. PHY-1203100 (USNA) and PHY-0754674 (FSU), as well as by the U.S. Department of energy, office of Nuclear of Nuclear Physics, under Contracts DE-AC02-06CH11357 (ANL) and DE-FG02-91ER40609 (Yale). One of us JFS-S would like to thank the Joyce Frances Adlard Cultural Fund for support. STNM acknowledges a postgraduate grant from the South African National Research Foundation and thanks the library staff of iThemba LABS for considerable help.